\documentclass[letterpaper,twocolumn,prb,aps,final,superscriptaddress]{revtex4-1}
\usepackage[latin9]{inputenc}
\usepackage{color}
\usepackage{textcomp}
\usepackage{amsmath}
\usepackage{amssymb}
\usepackage{graphicx}

\makeatletter

\pdfpageheight\paperheight
\pdfpagewidth\paperwidth

\providecommand{\tabularnewline}{\\}

\usepackage{pslatex}
\usepackage{listings}
\usepackage{xcolor}

\usepackage{verbatim}

\usepackage{fancyvrb}
\usepackage{color}
\usepackage{bm}

\makeatother

\begin{document}
\preprint{COMP/XIE2019}
\title{Generation of low-symmetry perovskite structures for \emph{ab initio}
computation}
\author{N. Xie}
\affiliation{School of Microelectronics \& State Key Laboratory for Mechanical
Behavior of Materials, Xi'an Jiaotong University, Xi'an 710049, China}
\author{J. Zhang}
\affiliation{School of Microelectronics \& State Key Laboratory for Mechanical
Behavior of Materials, Xi'an Jiaotong University, Xi'an 710049, China}
\author{N. Zhang}
\affiliation{Electronic Materials Research Laboratory--Key Laboratory of the Ministry
of Education and International Center for Dielectric Research, Xi'an
Jiaotong University, Xi'an 710049, China}
\author{X. Chen}
\affiliation{Department of Applied Physics, Aalto University, Espoo 00076, Finland}
\affiliation{BroadBit Batteries Oy, Espoo 02150, Finland}
\author{D. Wang}
\email{dawei.wang@xjtu.edu.cn}

\affiliation{School of Microelectronics \& State Key Laboratory for Mechanical
Behavior of Materials, Xi'an Jiaotong University, Xi'an 710049, China}
\date{\today}
\begin{abstract}
Ion displacements are the cause of the ferroelectricity in perovskites.
By properly shifting ions, \emph{ab initio} computations have been
extensively used to investigate properties of perovsites in various
structural phases. In addition to the relatively simple ion displacements,
perovskites have another type of structural distortion known as antiferrodistortion
or oxygen octahedron tilting. The interplay between these two types
of distortions have generated abundant structural phases that can
be tedious to prepare for \emph{ab initio} computation, especially
for large supercells. Here, we design and implement a computer program
to facilitate the generation of distorted perovskite structures, which
can be readily used for \emph{ab initio} computation to gain further
insight into the perovskite of a given structural phase.
\end{abstract}
\maketitle

\section{\label{sec:level1}Introduction}

Perovskites of the general formula ABX$_{3}$ had been used in many
areas due to their excellent physical properties\citep{application},
including dielectric, magnetic, electrical, optical, and catalytic
properties that have been widely investigated \citep{application-2,application-3,application-4}.
Their piezoelectricity and ferroelectricity have been used for ultrasonics,
transducers, pressure sensors, infrared thermoelectric detector, and
high density information storage for a long time and such applications
have gradually matured over time\citep{application-5}.\textbf{ }Recently,
it was also found that organic and inorganic (halide) perovskites
can be good candidates for solar cells and light emitting devices.\citep{Pcs}
In particular, perovskite solar cells have been intensively investigated
in search of greater conversion efficiency in addition to their simple
preparation process and low manufacturing cost.\citep{Pcs-2,Pcs-3,Pcs-4}

Predicting the stable phases of perovskites (and their stability)
remains a challenge in the investigation of related perovskites for
photovoltaics and electrocatalysts.\citep{Prediction,prediction-2}
All-inorganic perovskite materials {[}e.g., halide perovskites CsBX$_{3}$
(B = Sn and Pb; X = I, Br, and Cl){]}, without any volatile organic
components \citep{inorganic,inorganic-2,inorganic-3,inorganic-4},
exhibit superior thermal stability comparing to organic-inorganic
composite perovskite materials. However, understanding the stability
of their structural phases is also critical.\citep{instability} For
instance, cubic phase CsPbI$_{3}$ has a suitable band gap of 1.7\,eV
for high-efficiency photovoltaics, but it is not a thermodynamically
stable phase at room temperature.\citep{unstable,unstable-2} Thereby,
it is desirable to know the structural phase instabilities and be
able to predict the most stable phase and their basic properties can
be obtained via \emph{ab inito} computation before material synthesis.

Perovskites often exhibit a wide range of structural distortions originating
from lattice instabilities of their prototype cubic structure.\citep{distortion}
For instance, while a large number of perovskite ferroelectrics have
uniform polar distortions (e.g., in PbTiO$_{3}$, BaTiO$_{3}$, KNbO$_{3}$),
others can have more complex phases involving antiferroelectric ion
displacements and oxygen octahedron tilting (e.g., in PbZrO$_{3}$
and in SrTiO$_{3}$). The solid solution Pb(Zr$_{1-x}$Ti$_{x}$)O$_{3}$
embodied such complexity with its ever changing structural phases
with $x$. \citep{examples,PZT,PZT-2} Given the infinite possibilities
of perovskite compounds, reliably obtaining the most stable structural
phase for them is not a trivial work. The oxygen octahedron tilting
is particularly difficult to treat and analyze as the amplitudes and
energies due to the distortions are rather small.\citep{small}

In order to find the most stable phase, one common practice is to
calculate the the phonon spectra, obtain the eigenvectors of the unstable
phonon modes, shift the ions accordingly, and finally relax the constructed
structure using\emph{ ab initio} computation to obtain its energy.\citep{soft mode,soft mode-2,softmode-3,softmode-4,softmode-5}
In addition, the effective Hamiltonian approach has also been successful
in predicting and suggesting low-energy phases \citep{examples,localmode,localmode-2}.
This approach relies on the fact that low-energy phases are often
derived by ion displacements with respect to the original cubic cell
(with the $Pm\bar{3}m$ space group) and some distortion of the cell
shape. With the cubic phase as the reference structure, two sets of
parameters, which are (i) the local mode for ion displacements\citep{examples,dispersion,localmode,localmode-2,displacement,displacement-2}
and (ii) the antiferrodistortive mode for oxygen octahedron tiltings
\citep{small,tilt,tilt-2,tilt-3,tilt-4,tilt-5}, are used to describe
the distortion. The effective Hamiltonian approach uses both local
mode and local oxygen tilting as the dynamical variables to construct
the energy terms, which are the basis for the subsequent Monte-Carlo
and molecular dynamics simulations. The basic construct of the effective
Hamiltonian is the``local mode'' on each lattice site (often denoted
by $\boldsymbol{u}_{i}$), these local modes on all lattice sites
together can simulate the unstable optic phonon branches and reproduce
relevant physics \citep{dispersion}. With the local mode (and later
the oxygen octahedron tliting) and properly constructed energy terms,
Monte-Carlo or molecular dynamics simulations can predict the low
energy structures of a given perovskite\citep{energy,energy-1}. Afterward,
first principles calculations \citep{first} can be employed to verify
if a proposed structure is indeed the ground state by comparing it
to other structural phases. The comparison here requires the construction
of various crystal phases for \emph{ab initio} computation.

In constructing various structural phases, ion displacement is easy
to understand and keep track of. For instance, while BaTiO$_{3}$
has a complex sequence of ferroelectric phase transitions, the resulting
tetragonal, orthohombic, rohombohedral, ferroelectric phases can all
be understood in terms of the Ti ions shifting along different directions.
On the other hand, the second type of distortion (which could also
be part of the overall phase transition) involves small tiltings or
rotations of the BX$_{6}$ octahedra (usually the oxygen octahedra),
which is more difficult to deal with due to the corner sharing of
the X on neighboring sites that constrains possible movements. \citep{rotate,rotate-2,rotate-3}
For instance, the TiO$_{6}$ octahedrons in SrTiO$_{3}$ are not independent
of each other and can only exhibit certain patterns. One possibility
is that the TiO$_{6}$s rotate about the \textbf{c}-axis with neighboring
octahedra rotate in opposite directions. Some perovskites (e.g., CaTiO$_{3}$)
can demonstrate even more complex rotation patterns and it was M.
Glazer who first proposed a systematic notation to describe such patterns
\citep{tilt,tilt-2}, now commonly known as the Glazer notation.

In order to facilitate the search of stable structural phases with
first principles calculations, we design and implement a \texttt{Python}
program named \texttt{PyTilting} that can systematically generate
different structural phases that encompass distortions involving both
the local mode and the oxygen tilting. With this tool, it is possible
to automate the structure construction process, an essential step
in high throughput \emph{ab initio} computation \citep{calculation,calculation-2,calculation-3}.

\section{Structural distortions \label{sec:Structural-distortions}}

The perovskite structure contains a network of corner-sharing octahedra.
Considering such constraints, the following rules apply when we consider
tilings: (i) The axis of the tilting can be parallel to any crystallographic
axis, which requires a vector $\boldsymbol{\omega}=\left(\omega_{x},\omega_{y},\omega_{z}\right)$
to describe it; (ii) The amplitude of each tilting may be different
from the others, that is, $\omega_{x}$, $\omega_{y}$, and $\omega_{z}$
can be independent of each other; (iii) Two subsequent layers being
stacked along the tilting axis may be tilted in-phase or anti-phase.
However, neighboring octahedron on the plane perpendicular to the
tilting axis has be in anti-phase.

For a $2\times2\times2$ supercell with PBC, there are 23 tilting
patterns that can derive from the cubic perovskite.\citep{tilt,tilt-3}
However, when the displacement of atoms is involved and combined with
these tiltings, the situation becomes quite complicated. It is not
trivial in many cases to translate a desired distortion into a structure
setup with proper space group and symmetry, which can be readily employed
in \emph{ab initio} calculations. To address this challenge, we have
designed programs to realize such distortions in terms of ion displacements,
obtaining the low-symmetry structures from the cubic high-symmetry
structure ($Pm\bar{3}m$). In this section, we will first discuss
the general features of octahedron tilting and ion displacement.

\subsection{Octhedron tilting}

Glazer notation has become the standard way to describe tilting distortions,
concisely summarizing tiltlting patterns into strings such as $a^{0}b^{+}b^{+}$
and $a^{0}b^{+}b^{-}$. To decipher such strings, we first note that
they can be split into three groups, each describing the tilting situation
around one axis, in the order of \textbf{a},\textbf{b},\textbf{c}-axes.
Each group contains a letter specifying the magnitude of the tilting.
The second parameter is a superscript indicating whether the rotations
in adjacent layers are in the same or opposite directions ( in-phase
or out-of-phase rotation ). A negative superscript ($a^{-})$ indicates
that the tiltings of two neighboring octahedra, along the tilting
axis, are in the opposite directions, while a positive superscript
($a^{+})$ is used when they tilt in the same direction. A zero superscript
($a^{0})$ is used when no tilting occurs (see below for more illustration).
Given the fact that neighboring octahedron has to tilt/rotate oppositely
if the axis of rotation is perpendicular to the line connecting the
center of the two oxygen octahedra, each group also provides enough
information regarding the tiltings on each unit cell. This fact is
implicitly used in our program.

\begin{figure}[h]
\noindent \begin{centering}
\includegraphics[width=7cm]{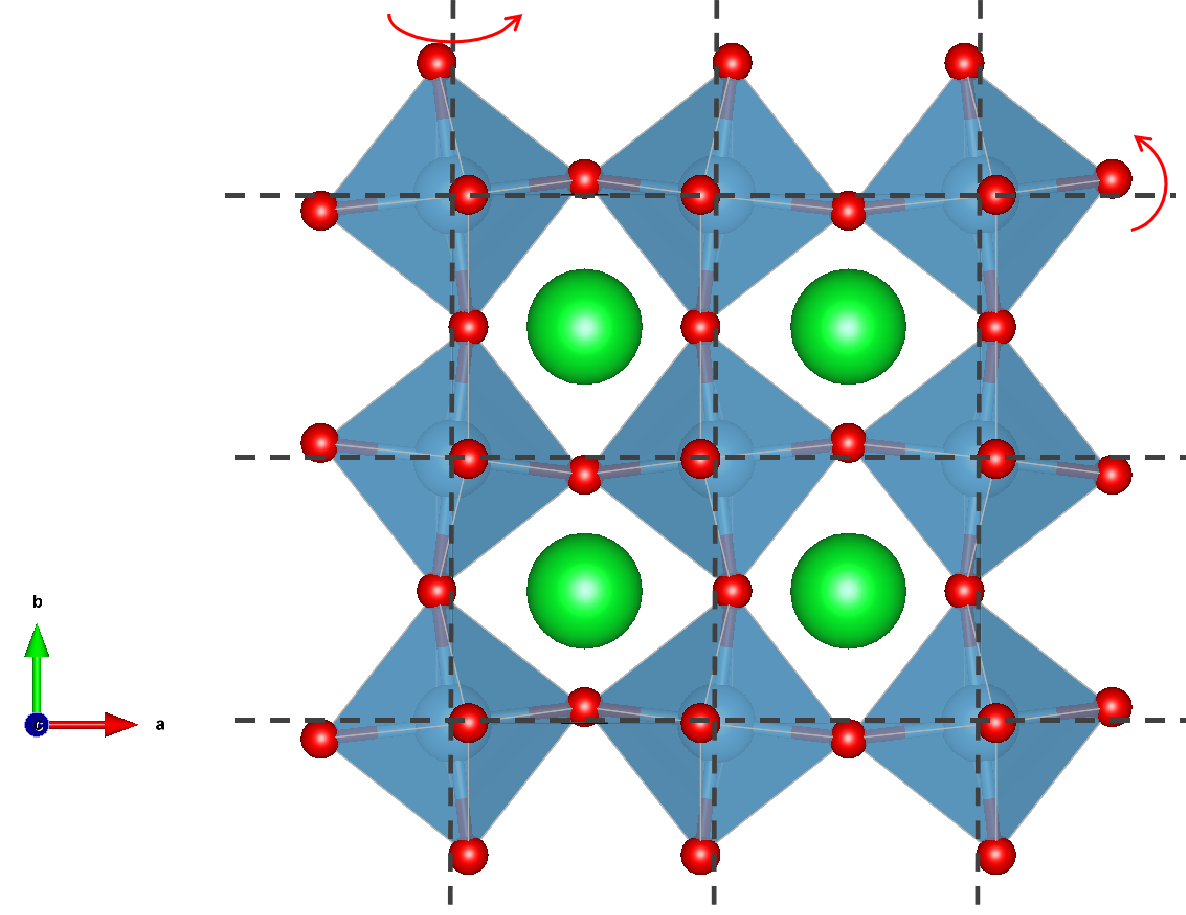}
\par\end{centering}
\noindent \centering{}\includegraphics[width=7cm]{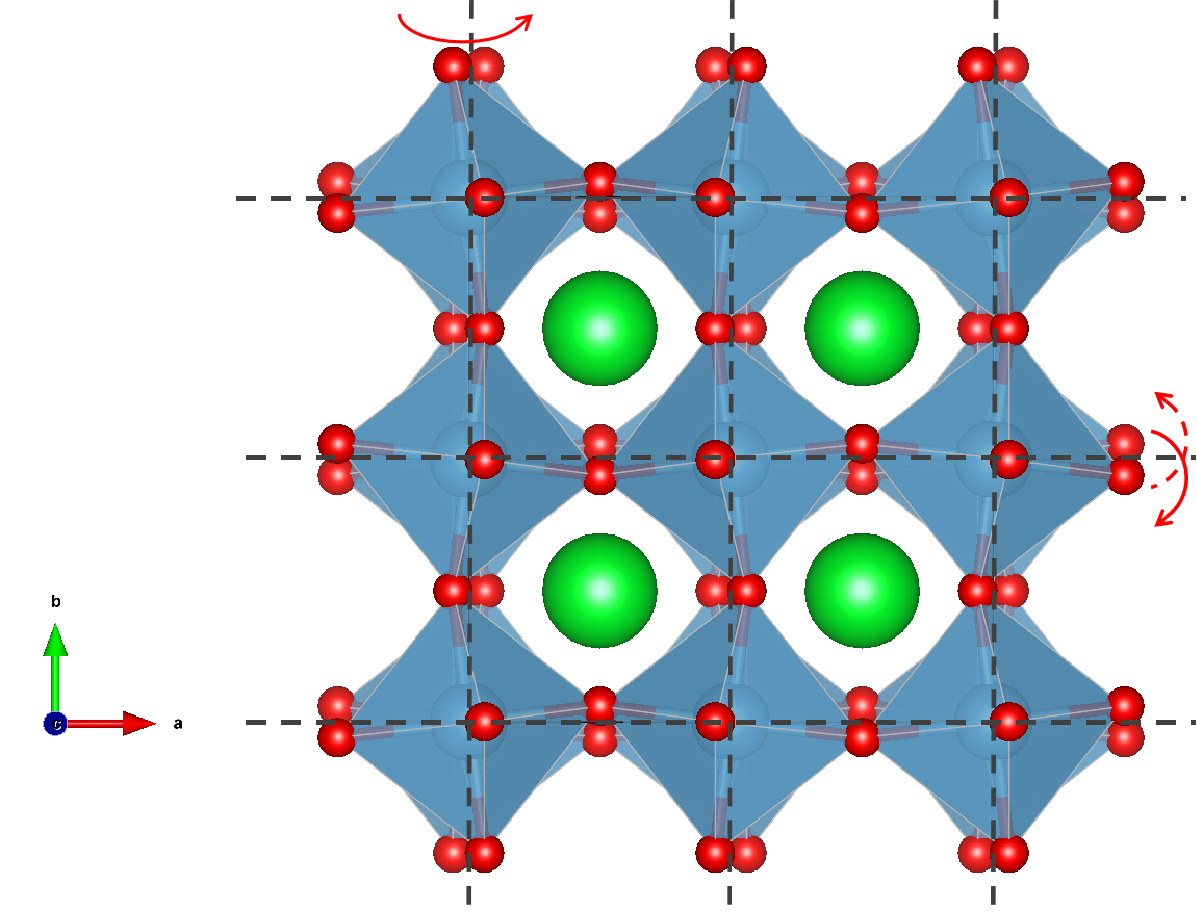}\caption{The projection of SrTiO$_{3}$ along the \textbf{c}-axis: (a) $a^{0}b^{+}b^{+}$and
(b) $a^{0}b^{+}b^{-}$ rotation patterns. The blue, red, and green
balls represent Ti, O, and Sr atoms, respectively. The gray grid line
provides a reference position for the Ti-O bond. The offset of the
oxygen atoms from the vertical grid lines indicates that the octahedron
rotates in the same direction around the \textbf{b}-axis, while the
zero offset from the horizontal grid lines indicates that the oxygen
octahedron does not rotate around the \textbf{a}-axis. In Figure (a),
the upper and lower oxygen octahedra (along the \textbf{c}-axis) rotate
in the same direction around the \textbf{c}-axis, while in (b) they
rotate in the opposite direction.\label{fig:example-of-glazor-notation}}
\end{figure}
 To illustrate the Glazer notation, we show the tilting patterns $a^{0}b^{+}b^{+}$
and $a^{0}b^{+}b^{-}$ in Fig. \ref{fig:example-of-glazor-notation}.
Clearly, these two Glazer notation specifies no tilting around the
\textbf{a}-axis, but tiltings around the \textbf{b}-axis and \textbf{c}-axis.
In addition, for both cases, counting along the \textbf{b}-axis, neighboring
unit cells have the same (in-phase) tiltings around the \textbf{b}-axis,
but counting along the \textbf{c}-axis, neighboring unit cells have
the opposite (out-of-phase) tiltings around the \textbf{c}-axis.

As we can see, it takes some effort to decipher the tilting patterns
specified by the Glazer notation. To reduce this unnecessary work,
our program fully understands the exact meaning of Glazer notation,
can decode them, and construct the desired structure for an arbitrary
(but properly chosen) supercells. For instance, it can easily reproduce
the 23 simple tilting systems with a $2\times2\times2$ supercell,
generating all the cases in which the oxygen octahedra have rotations
on one, two, or three axes. Large supercells with more exotic tilting
patterns can also be specified with minimal additional programming.

\subsection{Ion displacement\label{subsec:Ion-displacement}}

Having considered tilting, we now consider ion displacement, which
is easier because, unlike tilting, ion displacements on each unit
cell can be independently specified by three numbers $\boldsymbol{u}=\left(u_{x},u_{y},u_{z}\right)$
on each unit cell where $u_{x}$, $u_{y}$, and $u_{z}$ are the magnitudes
of ion shifting along the \textbf{a}-, \textbf{b}-, and \textbf{c}-axis.
With the information of the local mode, which specifies how the ions
inside each unit cell relatively shifts, the ion displacements can
all be specified throughout the system.

To be compatible with the periodic boundary condition, patterns of
the local mode over the the $2\times2\times2$ supercell can be specified
by using the high-symmetry points inside the Brillouin zone: $\Gamma=\left(0,0,0\right)$,
$X=(1,0,0)$, $M=(1,1,0)$, and $R=(1,1,1)$ (or other points with
larger supercells). For instance, the $\Gamma$ point specifies a
uniform $\boldsymbol{u}$ over the whole supercell. We note that the
local mode patterns for $u_{x}$, $u_{y}$, $u_{z}$ can be specified
independently. Practically, the local mode can be determined from
phonon eigenvectors (or eigenvectors of the force constant matrix)
\citep{localmode-2}. Here, for convenience, we usually just displace
the B atom to achieve the desired low-symmetry structure. Employing
more complex local mode is often inconsequential since the resulting
structure will be further relaxed by \emph{ab initio} softwares, most
of which can be set to respect the starting symmetry of a given system.

\section{Implementation}

With the above discussion in mind, we design a program \textsl{pyTilting}
to automatically generate the low-symmetry structures involving distortions
of both tilting and displacement. The program is designed to export
to ASE \texttt{atoms} object \citep{ASE} or a \texttt{cif} files,
which can be used in other \emph{ab initio} softwares, or directly
integrated with the specific software (e.g., GPAW \citep{GPAW-2}).
Furthermore, it is easily to download all the program and obtain related
documentation online\citep{pytilting}. While our implementation can
build an arbitrary supercell, we will focus on the $2\times2\times2$
supercell that can represent most interesting structures. In particular,
the 23 structures with only tilting can be represented with such a
supercell.

\begin{figure}[h]
\begin{centering}
\includegraphics[width=8.5cm]{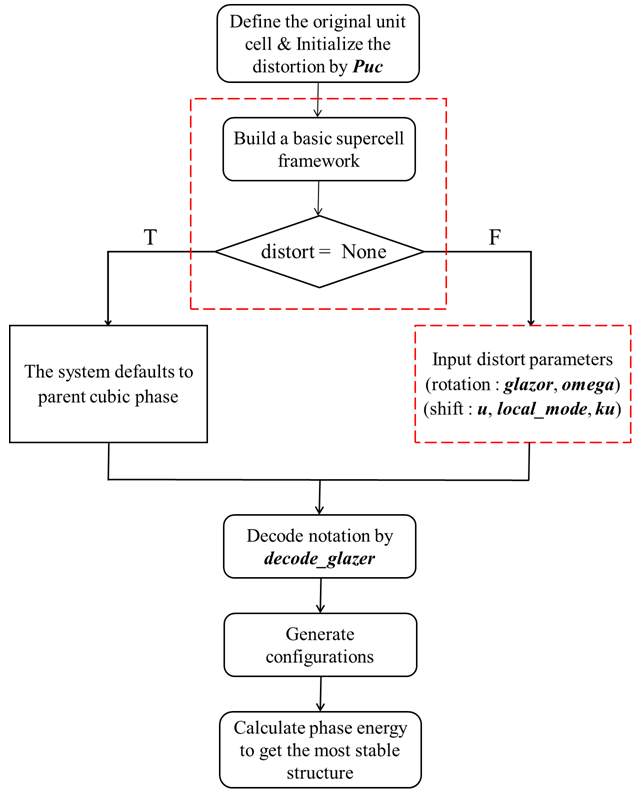}
\par\end{centering}
\noindent \caption{The execution flow of the program. \label{fig:The-execution-flow}}
\end{figure}
 Figure \ref{fig:The-execution-flow} shows the flow of the program.
It comprises several steps from top to bottom: (i) First, the initial
configuration is defined with the most basic parameters (e.g, atomic
symbols), setting the ions to their ideal cubic positions. All atoms
can shift but only oxygen atoms are allowed to rotate. A Python class
\texttt{Puc} is defined for this task; (ii) Next, the unit cell information
is inherited by the class \texttt{Distortion}, which builds a supercell
by repeating the unit cell according to the \texttt{grid} and \texttt{lattice\_constant}
parameter; (iii) In the next step, the program determines the structural
distortion given by the user. If no distortion is desired, the system
defaults to the parent phase ($Pm\bar{3}m$), otherwise the user's
input regarding the distortion is read; (iv) The Glazer notation is
decoded by the function class \texttt{decode\_glazer}. (v) Generating
the desired structures with the information regarding distortion;
(vi) Finally, the generated structures can be used in other \emph{ab
intio} calculations. The red boxes in Fig. \ref{fig:The-execution-flow}
indicate the necessary inputs to characterize the distortions. The
program and the inputs are explained below.

\subsection{Unit cell and supecell}

The program defines a Python class \texttt{Puc} that contains information
of perovskite primitive unit cell (PUC, each containing 5 atoms) and
basic operations, such as the \texttt{shift} and \texttt{rotate} functions
that operate on the PUC level. At initialization, the default atom
symbols are set to be ``A'', ``B'' and ``O'' referring to ABO$_{3}$,
but custom atom symbols can also be set. The five atoms in each PUC
are initially placed at their ideal positions with fractional coordinates,
$\boldsymbol{p}_{A}=[0.5,0.5,0.5]$, $\boldsymbol{p}_{B}=[0.0,0.0,0.0]$,
$\boldsymbol{p}_{O_{1}}=[0.5,0.0,0.0]$, $\boldsymbol{p}_{O_{2}}=[0.0,0.5,0.0]$,
$\boldsymbol{p}_{O_{3}}=[0.0,0.0,0.5]$.

Only oxygen atoms are allowed to rotate in the \texttt{rotate} function.
The new coordinates after rotation are obtained with $\boldsymbol{p}=c_{0}(\boldsymbol{\omega}\times\boldsymbol{p}_{0})+\boldsymbol{p}_{0}$,
where $\text{\ensuremath{\boldsymbol{\omega}}}$ is the vector of
angle that describes the tilting, $\boldsymbol{p}_{0}$ is the initial
position of the oxygen atom, and $c_{0}$ is a list represents the
$c/a$ ratio of the \textbf{c}-axis lattice constant divided by the
the \textbf{a}-axis lattice constant (defaults to 1). In the \texttt{shift}
the function, the coordinates of all atoms can be updated according
to the formula $\boldsymbol{p}=\Delta\boldsymbol{p}+\boldsymbol{p}_{0}$
where the displacement $\triangle\boldsymbol{p}$ is calculated according
to the local mode vectors and other input regarding dipole patterns.

After setting up the PUC, the \texttt{Distortion} class build a supercell
encompassing many identical PUCs and setup the distortion on the supercell
level, accommodating possible structural phases. The parameter \texttt{grid=$\left(n_{x},n_{y},n_{z}\right)$}
specifies how many unit cells repeat along the \textbf{a}-, \textbf{b}-,
and \textbf{c}-axis. 

The program includes some basic constructs consisting of several basic
parameters: (i) The atom symbols symbols for ABX$_{3}$; (ii) The
lattice constant; (iii) The \texttt{grid} parameter for supercell;
and (iv) The $c/a$ ratio. Other parameters specifying the actual
distortions and their use are discussed below.

\subsection{Specifying distortions}

Based the two distortion mechanisms discussed in Sec. \ref{sec:Structural-distortions},
tilting and ion displacement need to be specified as input for the
program.

\subsubsection{Decoding Glazer notation}

\begin{figure}[h]
\include{code_snippets/decoding.py}

\caption{Decoding Glazer notation.\label{fig:Decoding-Glazer-notation.}}
\end{figure}

For tilting, the program needs to properly change the positions of
oxygen atoms. Since the Glazer notation is the standard way to describe
the tilting in a perovskite system, it is natural to use it as the
input for the program and thus their meaning needs to be properly
decoded in the program. 

In a Glazer notation, the superscript $0$, $+$, or $-$ corresponds
to no tilting, in-phase, and out-of-phase tilting of neighboring oxygen
atoms, respectively. Such tilting patterns corresponds to high-symmetry
points in the Brillouin zone that are $\Gamma$ ($\boldsymbol{k}=[0,0,0]$),
$M$ ($\boldsymbol{k}=[0,1,1]$) and $R$ ($\boldsymbol{k}=[1,1,1]$).
Due to the constraint of the corner sharing of octahedra, the superscript
determines which point to choose. Let us use the tilting around the
\textbf{a}-axis as an example (i.e., $\omega_{x}$ in the first group
of a Glazer notation): (i) The superscript $-$ means $\boldsymbol{k}_{\omega_{x}}=\left[1,1,1\right]$
($R$), i.e., $\omega_{x}$ changes its sign whenever it moves to
the next nearest neighbor; (ii) The superscript $+$ means $\boldsymbol{k}_{\omega_{x}}=\left[1,1,1\right]$,
i.e., $\omega_{x}$ changes its sign whenever it moves to the next
first nearest neighbor along $y$ or $z$, but keep the same sign
when it moves along $x$; (iii) The superscript $0$ means $\boldsymbol{k}_{\omega_{x}}=\left[0,0,0\right]$,
i.e., the same tilting on all sites, which in fact forces $\omega_{x}=0$.
Similarly, the other two groups specifies how $\omega_{y}$ and $\omega_{z}$
are arranged over the whole crystal. 

The actual implementation to decode Glazer notation is shown in Fig.
\ref{fig:Decoding-Glazer-notation.}. Combining $\boldsymbol{k}_{\omega_{x}}$,
$\boldsymbol{k}_{\omega_{y}}$, and $\boldsymbol{k}_{\omega_{z}}$,
we know how the octahedron tilts in each unit cell. After determining
the arrangements of $\omega_{x,y,z}$, the program also checks if
the input tilting angles are consistent with the arrangement, and
if $\omega_{x}$, $\omega_{y}$ and $\omega_{z}$ shall be equal to
each other. For instance, the notation $a^{+}a^{+}c^{+}$, means $\omega_{x}=\omega_{y}\neq\omega_{z}$.
If inconsistency is encountered, the program will print warnings and
stop execution.

\subsubsection{Displacement}

\begin{figure}[h]
\include{code_snippets/shift.py}

\caption{Mathematical representation of the displacement.\label{fig:Displacement}}
\end{figure}

Two matrices are used to characterize ion displacements. The eigenvector
of the local mode (soft mode) in perovskites can be characterized
by four parameters: \texttt{dA} ($\xi_{A}$), \texttt{dB} ($\xi_{B}$),
\texttt{dO}$_{p}$ ($\xi_{O_{\perp}}$) and \texttt{dO}$_{q}$ ($\xi_{O_{\parallel}}$)
\citep{localmode-2}, which corresponds to displacements of the A
atom, B atom, the O atom forming the B-O bond along the displacement
direction, and the other two O atoms forming the B-O bonds perpendicular
to the displacement direction, respectively. Considering the fact
that the $x$, $y$, and $z$ directions are equivalent (for the cubic
phase used as the reference structure), the displacements can be mathematically
represented by a $5\times3$ matrix \texttt{dpq }as shown in Fig.
\ref{fig:Displacement}. The amplitude of the local mode is denoted
by $\boldsymbol{u}=\left(u_{x},u_{y},u_{z}\right)$, which contains
displacements along $x$ ($u_{x}$), $y$ ($u_{y}$), and $z$ ($u_{z}$)
directions represented by a $3\times3$ diagonal matrix \texttt{du}.
The actual displacement for each atom can be realized by the matrix
product of \texttt{dpq} and \texttt{du} as shown in Fig. \ref{fig:Displacement}. 

Similar to setting the patterns for tilting, we also need to set the
$\boldsymbol{k}$ vectors to specify how local mode \texttt{du} changes
from one unit cell to the next. With $k_{u_{x}}$, $k_{u_{y}}$, and
$k_{u_{z}}$ being independent of each other, for a $2\times2\times2$
supercell, $k_{u_{x}}$, $k_{u_{y}}$, and $k_{u_{z}}$ can be chosen
from $\Gamma=\left(0,0,0\right)$, $X=(1,0,0)$, $M=(1,1,0)$, or
$R=(1,1,1)$. 

\subsection{Example input}

To illustrate the use of the program, we show how to construct certain
structures of bismuth ferrite (BiFeO$_{3}$) step by step. It is known
that BiFeO$_{3}$ can have ferroelectric ($R3c$) and antiferroelectric
phases ($Pnma$) that, combined with its magnetic properties, make
it a popular multiferroic material.\citep{BFO,BFO-2,BFO-3,BFO-4}
The $R3c$ phase (space group: \#161) can be thought of as stretching
the ideal cubic structure in the {[}111{]} direction, which is made
more complex by the $a^{-}a^{-}a^{-}$ tilting, resulting in a rhombohedral
phase.\citep{BFO-3,BFO-4}

\begin{figure}[h]
\begin{centering}
\include{code_snippets/bfo_basic.py}
\par\end{centering}
\caption{Program input to setup basic properties of the material BiFeO$_{3}$.
\label{fig:tilting} }
\end{figure}
 The program first needs the lattice constant and atom symbols to
identify the material to be calculated. We set the lattice constant
of BiFeO$_{3}$ 3.96\,\AA, with a $2\times2\times2$ supercell (\texttt{grid}).
The parameter \texttt{covera} ($c/a$) accounts for the possible lattice
stretch along the \textbf{c} axis, which can be convenient sometimes
but not always necessary ($c/a$ defaults to 1.0). The setup at this
stage is shown in Fig. \ref{fig:tilting}.

\begin{figure}[h]
\begin{centering}
\include{code_snippets/bfo_distort.py}
\par\end{centering}
\caption{Program input to setup the distortion for the $R3c$ phase of BiFeO$_{3}$.
\label{fig:displacement}}
\end{figure}
 In the second stage, with the Glazer notation for BiFeO$_{3}$'s
$R3c$ phase being $a^{-}a^{-}a^{-}$ (\texttt{glazer}), we enter
the tilting angle $\boldsymbol{\omega}=$ (0.1,0.1,0.1) in radian,
which is just an estimate of the magnitude of tilting and completes
setup for the tilting distortion as shown in Fig. \ref{fig:tilting}.
We also need to endow a small displacement by input the magnitude
(\texttt{u}) and vector (\texttt{local\_mode}) of the local modes.
Assuming the amplitude of the local mode is given by $\boldsymbol{u}=\left(u_{x},u_{y},u_{z}\right)$
as mentioned in the Sec. \ref{subsec:Ion-displacement}, the program
automatically calculates and actuates the displacement for each ion
in a unit cell as shown in Fig. \ref{fig:Displacement}. Meanwhile,
the wave vector for the local mode is input as a matrix \texttt{k\_u
}with the rows representing $\boldsymbol{k}_{u_{x}}$, $\boldsymbol{k}_{u_{y}}$,
and $\boldsymbol{k}_{u_{z}}$. Knowing that the $R3c$ phase has polarization
along the $\left(111\right)$ direction, we set $\boldsymbol{u}=(0.1,0.1,0.1)$
and the wave vector at the $\Gamma$ point ($\boldsymbol{k}=[0,0,0]$
) as shown in Fig. \ref{fig:Displacement}.

\begin{figure}[h]
\begin{centering}
\includegraphics[width=7cm]{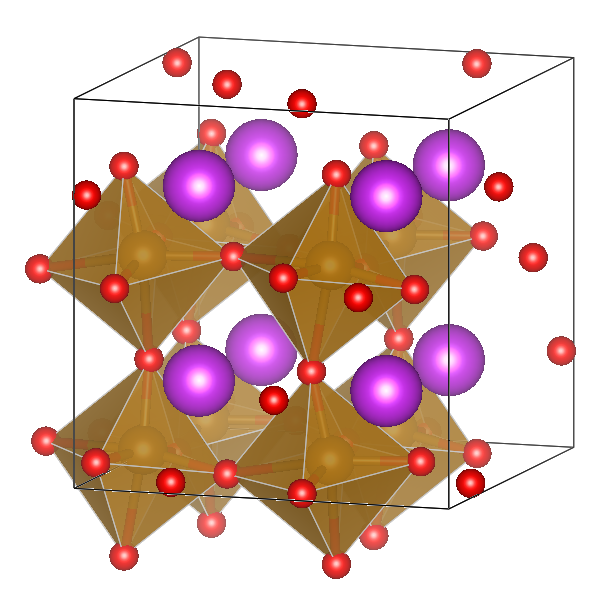}
\par\end{centering}
\caption{The generated $R3c$ phase of BiFeO$_{3}$ \label{fig:The-generated-BFO-R3c}.}

\end{figure}

Finally, we generate the atoms object that can be used directly by
ASE, GPAW, or exported to \texttt{cif} file. The symmetry of the generated
structure (see Fig. \ref{fig:The-generated-BFO-R3c}) can be checked
by ASE as shown in Fig. \ref{fig:The-generated-BFO-R3c}. 

\section{application \label{sec:application-to-halide}}

In this section, we apply the program to identify the most stable
phase of a halide perovskite, that is CsSnI$_{3}$. Halide perovskite
has attracted much attention for their use in solar cells for their
high energy conversion efficiency and tunable band gaps.\citep{CSI,CSI-2}
CsSnI$_{3}$ is an unusual perovskite that undergoes a series of complex
phase transitions and exhibits near-infrared emission at room temperature.
The phase-stabilized CsSnI$_{3}$ is important for its application
in the photovoltaic and optoelectronic fields\citep{CSI,CSI-3,CSI-4}. 

\subsection{Phonon analysis}

In order to reduce the number of possible structural phases, we perform
a screening to eliminate certain phases in order to save time and
computing resource by calculating the phonon band structure and perform
some initial analysis that focuses on the unstable modes. For instance,
negative (or imaginary) frequencies at the $\Gamma$ point are associated
with uniform global displacement of ions, which often implies ferroelectricity.

For ion displacement, in addition to the $\Gamma$ point, other points
from the Brillouin zone can also be used to specify the dipole patterns
(possible choices are constrained by the size of the supercell). On
the other hand, only two points are important for tilting, which are
(i) the $M$ point, $\boldsymbol{q}=(1/2,1/2,0)$ and (ii) the $R$
point, $\boldsymbol{q}=(1/2,1/2,1/2)$.

\begin{figure}[h]
\begin{centering}
\includegraphics[width=8.5cm]{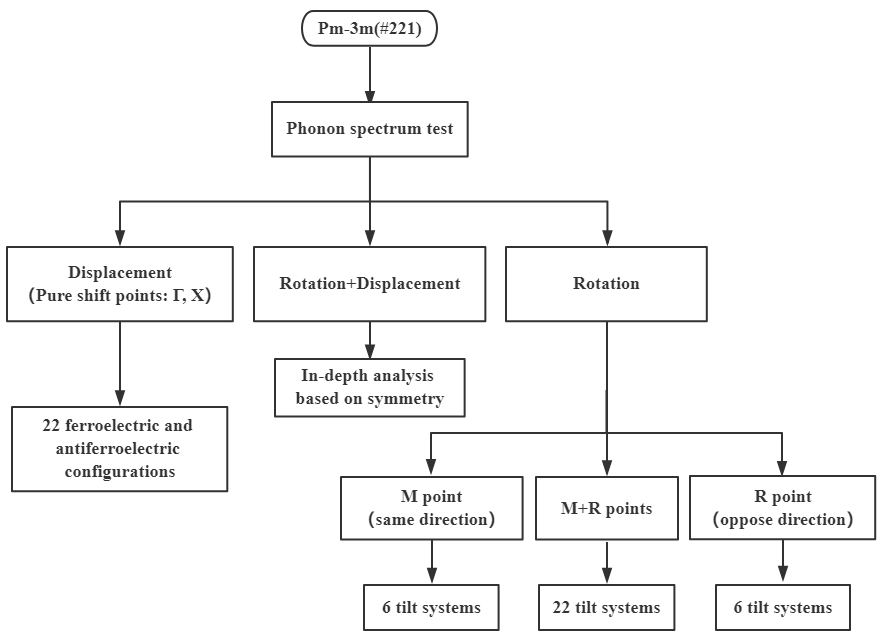}
\par\end{centering}
\noindent \caption{Combining phonon spectra to analyze phase transitions. \label{fig:Combining-phonon-spectra}}
\end{figure}
 The above analysis can be summarized into the flow chart in Fig.
\ref{fig:Combining-phonon-spectra}, where we assume the $M$ and
$R$ point are only related to tilting and the $\Gamma$ and $X$
point are related to displacements. The possibility that $M$ and
$R$ points accommodates antiferroelectric phases requires additional
analysis of the phonon eigenvectors at these two points. 

If the phonon spectrum only has a virtual frequency at the $\Gamma$,
the stable phase is most likely ferroelectric, as well as the $X$
point, other antiferroelectric phases also need to be considered --
there are 22 high-symmetry configurations for pure-shift points containing
14 with $u_{x}=u_{y}=u_{z}\neq0$, 6 with $u_{\alpha}=u_{\beta}\neq0$,
and and 2 with $u_{\alpha}\neq0$, where $\alpha,\beta=x,y,z$.

For pure tilting, there are three scenarios: (i) The virtual frequency
only appears at the $M$ point, resulting in 6 cases (out of 23) need
to be considered; (ii) The virtual frequency only appears at point
$R$, then again resulting in 6 cases (out of 23) need to be considered;
(iii) If both the $M$ and $R$ points have virtual frequencies, all
23 tilting patterns need to be considered. The most complicated scenario
occurs when all the high symmetry points have virtual frequencies,
resulting in the superposition of tilting and displacement that requires
further analysis.

\subsection{Results}

\begin{figure}[h]
\centering{}\includegraphics[width=8cm]{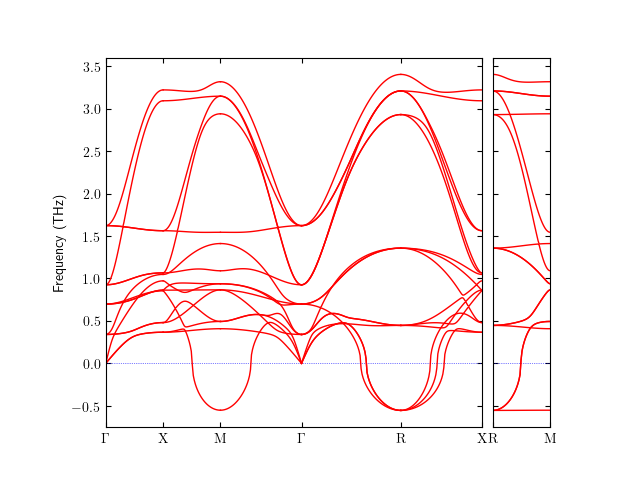}\caption{The phonon spectrum of CsSnI3\label{fig:The-phonon-spectrum}}
\end{figure}
 The phonon calculation for CsSnI$_{3}$ was done using Phonopy\citep{Phonony}
with GPAW. To get accurate results, we have used $2\times2\times2$
supercell in the phonon calculation, which agree with Ref. \citep{CSI-5}
very well. The phonon spectrum shown in Fig. \ref{fig:The-phonon-spectrum}
clearly demonstrates that CsSnI$_{3}$ has the strongest instabilities
at the $M$ and $R$ points, which involve all the tilting patterns.
On the other hand, there is no instability at the $\Gamma$ or $X$
point, which eliminates many possible structural phases.

To find the most stable structural phase, we use the program to generate
all the 23 tilting patters for \emph{ab initio} calculation using
the projector augmented plane wave (PAW) method as implemented in
the GPAW Package\citep{GPAW,GPAW-2} with the Perdew\textminus Burke\textminus Ernzerhof
(PBE) functional\citep{PBE}. A $4\times4\times2$ Monkhorst-Pack\citep{M-P}
sampling is used for the $k$-space integration. The valence orbitals
used in the first-principles calculations are: Cs (5$s$ 6$s$ 5$p$),
Sn (5$s$ 5$p$ 4$d$ ) and I (5$s$ 5$p$). The structures are relaxed
with a plane-wave cutoff of 900 eV until atomic forces fall below
0.05 eV/\AA. As shown in Figure 7, it is clear that the highest energy
belongs to the parent phase of $Pm\bar{3}m$, while the phase with
the lowest energy has the tilting $a^{+}b^{-}b^{-}$ .

\begin{table}[h]
\begin{centering}
\begin{tabular}{|c|c|c|c|}
\hline 
Number & Tilting & Symmetry & Energy(meV)\tabularnewline
\hline 
\hline 
3 tilts &  &  & \tabularnewline
\hline 
1 & $a^{+}b^{+}c^{+}$ & $Immm$ (\#71) & -83.7\tabularnewline
\hline 
2 & $a^{+}b^{+}b^{+}$ & $Immm$ (\#71) & -83.0\tabularnewline
\hline 
3 & $a^{+}a^{+}a^{+}$ & $Im\bar{3}$ (\#204) & -82.5\tabularnewline
\hline 
4 & $a^{+}b^{+}c^{-}$ & $Pmmn$ (\#59) & -130.2\tabularnewline
\hline 
5 & $a^{+}a^{+}c^{-}$ & $P4_{2}/nmc$ (\#137) & -118.6\tabularnewline
\hline 
6 & $a^{+}b^{+}b^{-}$ & $Pmmn$ (\#59) & -124.9\tabularnewline
\hline 
7 & $a^{+}a^{+}a^{-}$ & $P4_{2}/nmc$ (\#137) & -118.9\tabularnewline
\hline 
8 & $a^{+}b^{-}c^{-}$ & $P2_{1}/m$ (\#11) & -150.4\tabularnewline
\hline 
9 & $a^{+}a^{-}c^{-}$ & $P2_{1}/m$ (\#11) & -144.2\tabularnewline
\hline 
10 & $a^{+}b^{-}b^{-}$ & $Pnma$ (\#62) & -151.2\tabularnewline
\hline 
11 & $a^{+}a^{-}a^{-}$ & $Pnma$ (\#62) & -148.8\tabularnewline
\hline 
12 & $a^{-}b^{-}c^{-}$ & $P\bar{1}$ (\#2) & -97.8\tabularnewline
\hline 
13 & $a^{-}b^{-}b^{-}$ & $C2/c$ (\#15) & -92.6\tabularnewline
\hline 
14 & $a^{-}a^{-}a^{-}$ & $R\bar{3}c$ (\#167) & -91.4\tabularnewline
\hline 
2 tilts &  &  & \tabularnewline
\hline 
15 & $a^{0}b^{+}c^{+}$ & $Immm$ (\#71) & -91.6\tabularnewline
\hline 
16 & $a^{0}b^{+}b^{+}$ & $I4/mmm$ (\#139) & -88.8\tabularnewline
\hline 
17 & $a^{0}b^{+}c^{-}$ & $Cmcm$ (\#63) & -124.6\tabularnewline
\hline 
18 & $a^{0}b^{+}b^{-}$ & $Cmcm$ (\#63) & -122.1\tabularnewline
\hline 
19 & $a^{0}b^{-}c^{-}$ & $C2/m$ (\#12) & -103.9\tabularnewline
\hline 
20 & $a^{0}b^{-}b^{-}$ & $Imma$ (\#74) & -100.8\tabularnewline
\hline 
1 tilt &  &  & \tabularnewline
\hline 
21 & $a^{0}a^{0}c^{+}$ & $P4/mbm$ (\#127) & -82.9\tabularnewline
\hline 
22 & $a^{0}a^{0}c^{-}$ & $I4/mcm$ (\#140) & -83.9\tabularnewline
\hline 
no tilt &  &  & \tabularnewline
\hline 
23 & $a^{0}a^{0}a^{0}$ & $Pm\bar{3}m$ (\#221) & 0.0\tabularnewline
\hline 
\end{tabular}
\par\end{centering}
\caption{23 tilting systems for CsSnI$_{3}$ and their energies (the energy
of the $Pm\bar{3}m$ phase is used as the reference energy). \label{tab:23-tilting-systems}}
\end{table}

Table \ref{tab:23-tilting-systems} shows all the 23 tilting patterns
for CsSnI$_{3}$ with its Glazer notation and symmetry group. We note
the results are consistent with the tilting system proposed by Glazer,
except for the tilting systems No. 5 ($a^{+}b^{+}b^{-}$) and No.7
($a^{+}a^{+}a^{-}$). These two systems should have the $P42/nmc$
phase (space group: 137) instead of the $Pmmn$ phase (space group:
59) that was initially proposed \citep{tilt}, which has been confirmed
by other literature.\citep{small,tilt-3,tilt-5,137}.

The energy of the four systems No.8 to No.11 is quite close to each
other, with a difference of \textasciitilde 2.5 meV. Coincidentally,
this phenomenon also appears for No.1 to No.3, and their energy difference
is as small as \textasciitilde 1 meV, and No. 8 and No. 10 with difference
less than 1\,meV. It is also worth mentioning that even if some configurations
have the same space group, their tilting patterns are still different
(see No.1, 2 and 15 in Tab. \ref{tab:23-tilting-systems}), resulting
in different energies.

\begin{figure}
\begin{centering}
\includegraphics[width=8cm]{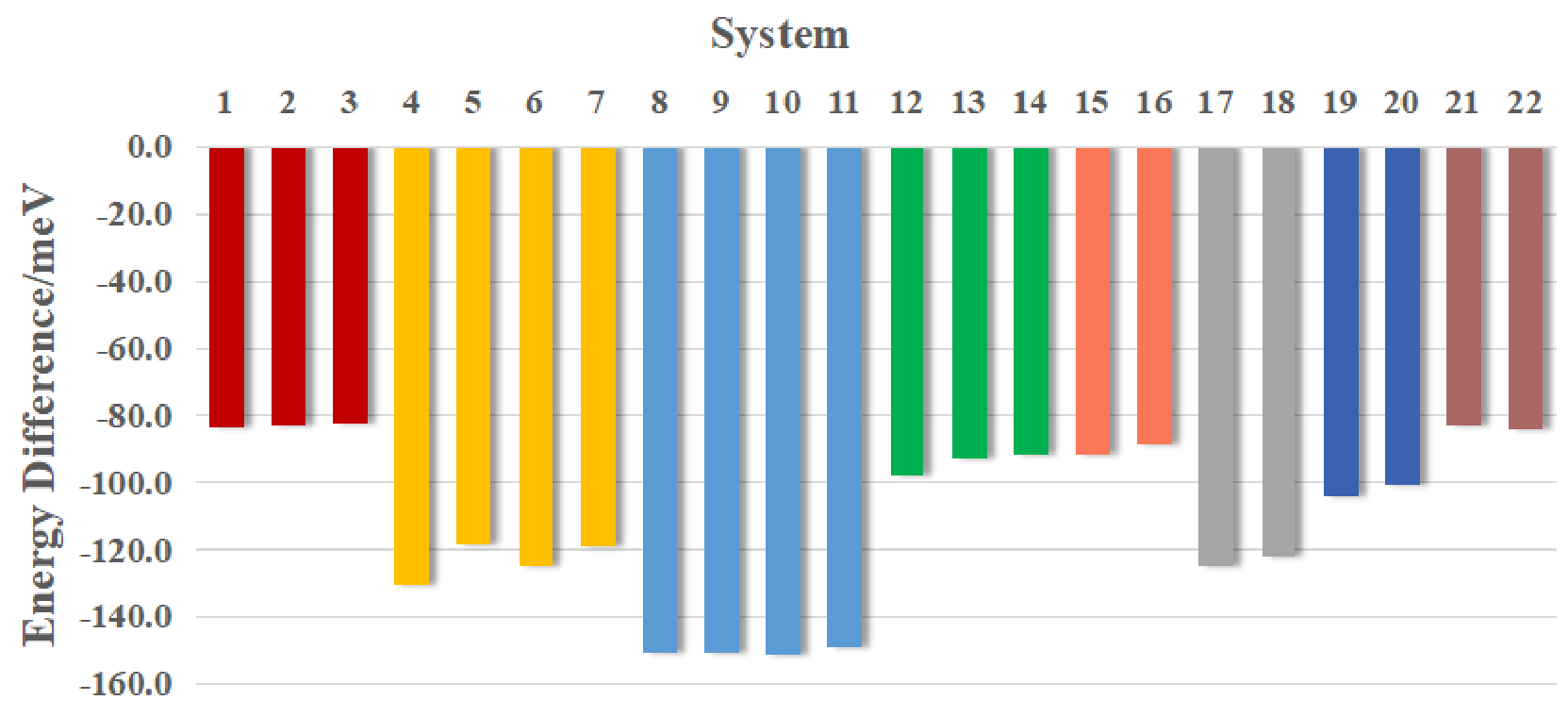}
\par\end{centering}
\caption{The energy of the 22 tilting configurations after geometry relaxation.
\label{fig:Energy-of-23}}
\end{figure}

\section{Discussion}

\subsection{CsSnI$_{3}$}

The space groups and the most stable structure predicted here agree
well with experimental results and other theoretical or computational
results \citep{CSI,CSI-2,CSI-3,CSI-4,CSI-5}. For CsSnI$_{3}$ crystal
(black phase), it is known to undergo a number of successive phase
transitions as a function of temperature, corresponding to rotations
and rearrangements of the SnI $_{6}$ octahedron. The cubic phase
(B-$\alpha$) is only stable at high temperature, which turns into
the tetragonal $P4/mbm$ phase at 380 K (B-$\beta$), then to the
orthorhombic $Pnma$ at 300 K (B-$\gamma$).

The tolerance factor $t$ is a powerful predictor to tilting distortions.
It turns out that the case size for the Cs atom in the ideal cubic
phase , which is set by the Sn-I octahedron size, is slightly too
large. By the definition given by Benedek and Fennie\citep{B-F},
$t=R_{AC}/\sqrt{2}R_{BC}$ with $R_{AC}$ and $R_{BC}$ being the
ideal bond lengths. Using the ideal bond lengths based on Shannon
ionic radii \citep{iion}, we find $t=0.90$ for CsSnI$_{3}$, indicating
that tilting can suppress ferroelectricity, consistent with the general
trend of adopting the $Pnma$ phase with $t<1$. The driving force
for this stabilization is to optimize the environment for the A cation
(Cs atom here).\citep{CSI-5} Moreover, the $\alpha\rightarrow\beta$
phase transition corresponds to the Glazer tilting $a^{0}a^{0}c^{+}$,
and the $\gamma$ phase icharacterized by both octahedral twists about
the c axis and tilts about the tetragonal a axis, which refers to
the Glazer notation $a^{+}b^{-}b^{-}$. Our results confirm the stability
of the $Pnma$ being the most stable phase.

\subsection{Tilting}

Our results indicate that the tilting patterns can be divided into
several groups, reflecting their  geometry features: (1)$a^{+}b^{+}c^{+}$
(No.1\textasciitilde 3); (2) $a^{+}b^{+}c^{-}$ ( No.4\textasciitilde 7);
(3) $a^{+}b^{-}c^{-}$ (No.8\textasciitilde 11); (4) $a^{-}b^{-}c^{-}$
(No.12\textasciitilde 14); (5) $a^{0}b^{+}c^{+}$ (No.15, 16); (6)
$a^{0}b^{+}c^{-}$ (No.17, 18); (7) $a^{0}b^{-}c^{-}$ (No.19, 20);
(8) $a^{0}a^{0}c^{+}$ (system No.21) and (9) $a^{0}a^{0}c^{-}$ (system
No.22). The tilting patterns of the same group tend to have extremely
close energy, as demonstrated in Fig. \ref{fig:Energy-of-23} because
their mutual conversion only requires a certain adjustment of the
tilting angle rather than a change of the the tilting patter, while
such adjustment can be actuated in \emph{ab initio} calculations.
For instance, $a^{+}b^{-}c^{-}$ could become very close to $a^{+}b^{-}b^{-}$,
$a^{0}b^{+}c^{+}$ is similar to $a^{0}b^{+}b^{+}$. This is likely
because, after relaxation, the tilting angles become more similar
to each other, moving towards the high-symmetry configurations. 

According to our results, in each group, if two (or three) axes have
the same tilting angle, then higher symmetry is achieved, but resulting
in increased energies. In addition, the $R$-point-related tilting
patterns seem to have lower energy than $M$-point-related tilting
patterns when other factors being equal (e.g. system No.1 and 12,
No.2 and 10, No.21 and 22). The calculation of the 23 tilt systems
shows that the energy of the structure depends mainly on tilting patterns,
not the actual tilitng angles.\textcolor{red}{{} }

Finally, it should be mentioned that, for some tilting patterns, small
distortions (in addition to tilting) of the octahedra must occur to
preserve the connectivity of the octahedra. Such a requirement results
in symmetries (and space groups) that are different from expectations
purely from the consideration of tilting. For instance, the four tilting
patterns in group 2 (No. 4\textasciitilde 7) was initially proposed
to produce an orthogonal phase (with the space group of 59) \citep{tilt}.
However, later research\citep{small,tilt-3,tilt-5,137} examined the
geometrical constraints and found that two of them have the $P4_{2}/nmc$
symmetry, requiring extra distortion to the octahedra. In other words,
such tilting patterns cannot be accommodated without changing the
B--O bond lengths. Due to this type of additional distortion, tilting
patterns No.5 and 7 are able to form a more symmetrical tetragonal
phase, and resulting in higher energies than No.4 and 6, respectively.

\section{Summary}

In this work, we have developed a program to generate various tiliting
patterns for perovskites with the ability to set up ion displacments
as well. This program can generate desired configurations for arbitrary
superlattices, which is useful to deal with complex structural phases
or complex perovskites made by doping or solid solution. 

The program adopts two strategies to deal with two major distortions
with perovskites: (1) For the tilting, we derive 23 tilting patterns
based on the Glazer notation; (2) Use local-mode-based displacements
to form polar phases. Combining these two strategies enables us to
generate hybrid structural phases made of both tiling and ion displacement. 

To test the program, we have taken CsSnI$_{3}$ as an example and
use the program to find its most stable structural phase. The strategy
is to first obtain the phonon spectrum in order to reduce the number
of different phases to examine. With the aid of our software package,
the 23 tilting systems are built and their energies calculated. It
is found that the $Pnma$ is the most stable phase, corresponding
to the $a^{+}b^{-}b^{-}$ tilting with its energy reduced by \textasciitilde 150\,meV
from the cubic $Pm\bar{3}m$ phase.

The computer program is written in Python, which can directly generate
configuration files, or easily be integrated into other software packages
for first-principles calculations. It is open source online currently
and able to download and use. Since our program can build arbitrary
supercells, doping effects on a given structure can be dealt with
by replacing certain ions in the system after the desired structural
phases are generated. With such a tool available, we can build a vast
number of different structural phases, the next step will be finding
effective ways to screen those candidates quickly so that first-principles
computations will be able to quickly provide us the most stable phase
of a given perovksite.
\begin{acknowledgments}
This work is financially supported by the National Natural Science
Foundation of China, Grant No. 11574246, U1537210, and 11974268. X.C.
Thanks the financial support from Academy of Finland Projects 308647
and European Union's Horizon2020, Grant No. 760930. X.C. and D.W.
thank the support form CSC (IT Center for Science, Finland), project
2001447, for providing computation resource. D.W. also thanks the
support from the Chinese Scholarship Council (201706285020).
\end{acknowledgments}


\begin{thebibliography}{99}
\bibitem{application}M. E. Lines, and A. M. Glass, \textit{Principles
and Applications of Ferroelectrics and Related Materials }(Clarendon
Press, Oxford, 1977).

\bibitem{application-2}E.A.R. Assirey, Saudi Pharmaceutical Journal
\textbf{27} (2019), 817. https://doi.org/10.1016/j.jsps.2019.05.003. 

\bibitem{application-3}N. A. Spaldin, and R. Ramesh, Nat. Mater.
\textbf{18} (2019), 203. https://doi.org/10.1038/s41563-018-0275-2

\bibitem{application-4}D. Wang, A. A. Bokov, Z.-G. Ye, J. Hlinka,
and L. Bellaiche, Nat. Commun. \textbf{7} (2016), 11014. https://doi.org/10.1038/ncomms11014

\bibitem{application-5}Z.-G. Ye, Materials Today \textbf{11} (2008),
70. https://doi.org/10.1016/S1369-7021(08)70258-X

\bibitem{Pcs} L. N. Quan, F. P. García de Arquer, R. P. Sabatini,
and E. H. Sargent, Adv. Mater. \textbf{30} (2018), 1801996. https://doi.org/10.1002/adma.201801996

\bibitem{Pcs-2}N. J. Jeon, J. H. Noh, Y. C. Kim, W. S. Yang, S. Ryu,
and S. I. Seok, Nature. \textbf{517} (2015), 897. https://doi.org/10.1038/nature14133

\bibitem{Pcs-3}H. Chen, F. Ye, W. Tang, J. He, M. Yin, Y. Wang, F.
Xie, E. Bi, X. Yang, and M. Grätzel, Nature \textbf{550} (2017), 92.
https://doi.org/10.1038/nature23877

\bibitem{Pcs-4}K. Aitola, K. Domanski, J. P. Correa-Baena, K. Sveinbjornsson,
M. Saliba, A. Abate, M. Gratzel, E. Kauppinen, E. M. J. Johansson,
W. Tress, A. Hagfeldt, and G. Boschloo, Adv. Mater. \textbf{29} (2017),
1606398. https://doi.org/10.1002/adma.201606398

\bibitem{Prediction}S. M. Woodley, and R. Catlow, Nat. Mater. \textbf{7
}(2008), 937. https://doi.org/10.1038/nmat2321

\bibitem{prediction-2}C. J. Bartel, C. Sutton, B. R. Goldsmith, et
al. Sci. Adv. \textbf{5} (2019), eaav0693. https://doi.org/10.1126/sciadv.aav0693.

\bibitem{inorganic}M. Saliba, T. Matsui, K. Domanski, J. Y. Seo,
A. Ummadisingu, S. M. Zakeeruddin, J. P. Correa-Baena, W. R. Tress,
A. Abate, A. Hagfeldt, and M. Gratzel, Science \textbf{354} (2016),
206. https://doi.org/10.1126/science.aah5557

\bibitem{inorganic-2}J. Liang, C. X. Wang, Y. R. Wang, Z. R. Xu,
Z. P. Lu, Y. Ma, H. F. Zhu, Y. Hu, C. C. Xiao, X. Yi, G. Y. Zhu, H.
L. Lv, L. B. Ma, T. Chen, Z. X. Tie, Z. Jin, and J. Liu, J. Am. Chem.
Soc. \textbf{138} (2016), 15829. https://doi.org/10.1021/jacs.6b10227

\bibitem{inorganic-3}A. Swarnkar, A. R. Marshall, E. M. Sanehira,
B. D. Chernomordik, D. T. Moore, J. A. Christians, T. Chakrabarti,
and J. M. Luther, Science \textbf{354} (2016), 92. https://doi.org/10.1126/science.aag2700

\bibitem{inorganic-4}M. Kulbak, S. Gupta, N. Kedem, I. Levine, T.
Bendikov, G. Hodes, and D. Cahen, J. Phys. Chem. Lett. \textbf{7}
(2016), 167. https://doi.org/10.1021/acs.jpclett.5b02597

\bibitem{instability} R. Wang, M. Mujahid, Y. Duan, Z.-K. Wang, J.
Xue, and Y. Yang, Adv. Funct. Mater. (2019), 1808843. https://doi.org/10.1002/adfm.201808843.

\bibitem{unstable} J. Sun, S. Huang, J. Hu, et al. J. Am. Chem. Soc.
\textbf{140} (2018), 11705. https://doi.org/10.1021/jacs.8b05949

\bibitem{unstable-2} J. B. Hoffman, A. L. Schleper, and P. V. Kamat,
J. Am. Chem. Soc. \textbf{138} (2016), https://doi.org/8603.10.1021/jacs.6b04661

\bibitem{distortion}M. E. Lines, A. M. Glass, and G. Burns, Physics
Today, \textbf{31} (1978), 56. https://doi.org/10.1063/1.2995188

\bibitem{examples}U. V. Waghmare, and K. M. Rabe, Phys. Rev. B \textbf{55}
(1997), 6161. https://doi.org/10.1103/PhysRevB.55.6161

\bibitem{PZT} Igor A. Kornev, L. Bellaiche, P.-E. Janolin, B. Dkhil,
and E. Suard. Phys. Rev. Lett. \textbf{97} (2006), 157601. https://doi.org/10.1103/PhysRevLett.97.157601

\bibitem{PZT-2}N. Zhang, H. Yokota, A. M. Glazer, Z. Ren, D. A. Keen,
D. S. Keeble, P. A. Thomas and Z.-G. Ye. Nat. Commun. \textbf{5} (2014),
5231. https://doi.org/10.1038/ncomms6231

\bibitem{small}M. A. Islam, J. M. Rondinelli, and J. E. Spanier,
J. Phys.: Condens. Matter \textbf{25} (2013), 175902 https://doi.org/10.1088/0953-8984/25/27/275902

\bibitem{soft mode}M. Mori, and H. Saito, J. Phys. C: Solid State
Phys. \textbf{19} (1986), 2391. https://doi.org/10.1088/0022-3719/19/14/005

\bibitem{soft mode-2}K. Gesi, J. D. Axe, and G. Shirane, Phys. Rev.
B \textbf{5} (1972), 1933. https://doi.org/10.1103/PhysRevB.5.1933

\bibitem{softmode-3}J. F. Scott, Rev. Mod. Phys. \textbf{46} (1974),
83. https://doi.org/10.1103/revmodphys.46.83

\bibitem{softmode-4}P. M. Woodward, Acta Cryst.: Sect. B \textbf{53}
(1997), 44. https://doi.org/10.1107/S0108768196012050

\bibitem{softmode-5}Y. Fujii, and S. Hoshino, Phys. Rev. B \textbf{9}
(1974), 4549. https://doi.org/10.1103/PhysRevB.9.4549

\bibitem{dispersion}Ph. Ghosez, and K. M. Rabe. Phys. Rev. B \textbf{60}
(1999), 836. https://doi.org/10.1103/PhysRevB.60.836

\bibitem{energy}D. Wang, J. Hlinka, A.A. Bokov, Z.-G. Ye, P. Ondrejkovic,
J. Petzelt, and L. Bellaiche, Nat. Commun. \textbf{5} (2014), 5100.
https://doi.org/10.1038/ncomms6100

\bibitem{energy-1}Z. Jiang, B. Xu, F. Li, D. Wang, and C.-L. Jia,
Phys. Rev. B \textbf{91} (2015), 014105. https://doi.org/10.1103/PhysRevB.91.014105

\bibitem{first}D. S. Sholl and J. A. Steckel, (2010), \textit{Density
Functional Theory:A Practical Introduction.}(Wiley, 2009). https://doi.org/10.1002/9780470447710

\bibitem{localmode}W. Zhong, D. Vanderbilt, and K. M. Rabe, Phys.
Rev. Lett. \textbf{73} (1994), 1861. https://doi.org/10.1103/PhysRevLett.73.1861

\bibitem{localmode-2}W. Zhong, D. Vanderbilt, and K. M. Rabe.\textit{
}Phys. Rev. B \textbf{52} (1995), 6301. https://doi.org/10.1103/PhysRevB.52.6301

\bibitem{displacement}K. M. Rabe, and U. V. Waghmare, Ferroelectrics
\textbf{164} (1995), 15. https://doi.org/10.1080/00150199508221827

\bibitem{displacement-2}W. Zhong, and D. Vanderbilt, Phys. Rev. Lett.
\textbf{74} (1995), 2587. https://doi.org/10.1103/PhysRevLett.74.2587

\bibitem{tilt}A. M. Glazer, Acta Cryst.: Sect. B \textbf{28} (1972),
3384. https://doi.org/10.1107/S0567740872007976

\bibitem{tilt-2}A. M. Glazer, Acta Cryst.: Sect. A \textbf{31} (1975),
756. https://doi.org/10.1107/S0567739475001635

\bibitem{tilt-3}P. M. Woodward, Acta Cryst. B \textbf{53} (1997),
32. https://doi.org/10.1107/S0108768196010713

\bibitem{tilt-4}R. J. Angel, J. Zhao, and N. L. Ross, Phys. Rev.
Lett. \textbf{95} (2005), 025503. https://doi.org/10.1103/PhysRevLett.95.025503

\bibitem{tilt-5} C. J. Howard, and H. T. Stokes, Acta Cryst.: Sect.
B \textbf{54} (1998), 782. https://doi.org/10.1107/S010876810200890X

\bibitem{rotate}H. T. Stokes, E. H. Kisi, D. M. Hatch, and C. J.
Howard, Acta Cryst.: Sect. B \textbf{58} (2002), 934. https://doi.org/10.1107/S0108768102015756

\bibitem{rotate-2}C. N. W. Darlington, Acta Cryst.: Sect. A \textbf{58}
(2002), 299. https://doi.org/10.1107/S0108767301016579

\bibitem{rotate-3}C. J. Howard, and H. T. Stokes, Acta Cryst.: Sect.
A \textbf{61} (2005), 93. https://doi.org/10.1107/S0108767304024493

\bibitem{calculation}K. F. Garrity, Phys. Rev. B \textbf{97} (2018),
024115. https://doi.org/10.1103/PhysRevB.97.024115

\bibitem{calculation-2}A. Jain, G. Hautier, C. J. Moore, S. P. Ong,
C. C. Fischer, T. Mueller, K. A. Persson, and G. Ceder, Comput. Mater.
Sci. \textbf{50} (2011), 2295. https://doi.org/10.1016/j.commatsci.2011.02.023

\bibitem{calculation-3}S. Curtarolo, W. Setyawan, S. Wang, J. Xue,
K. Yang, R. H. Taylor, L. J. Nelson, G. L. Hart, S. Sanvito, M. Buongiorno
Nardelli et al., Comput. Mater. Sci. \textbf{58} (2012), 227. https://doi.org/10.1016/j.commatsci.2012.02.002

\bibitem{ASE} A. H. Larsen, J.J. Mortensen, J. Blomqvist, I.E. Castelli,
R. Christensen, M. Du\l ak, J. Friis, M.N. Groves, B. Hammer, C. Hargus,
E.D. Hermes, P.C. Jennings, P.B. Jensen, J. Kermode, J.R. Kitchin,
E.L. Kolsbjerg, J. Kubal, K. Kaasbjerg, S. Lysgaard, J.B. Maronsson,
T. Maxson, T. Olsen, L. Pastewka, A. Peterson, C. Rostgaard, J. Schiøtz,
O. Schütt, M. Strange, K.S. Thygesen, T. Vegge, L. Vilhelmsen, M.
Walter, Z. Zeng, and K.W. Jacobsen, \textit{The atomic simulation
environment---a Python library for working with atoms}, J. Phys.:
Condens. Matter \textbf{29}, 273002 (2017). https://doi.org/10.1088/1361-648X/aa680e

\bibitem{BFO}C. Michel, J. M. Moreau, G. D. Achenbach, R. Gerson,
and W. J. James, Solid State Commun. \textbf{7} (1969). https://doi.org/701.10.1016/0038-1098(69)90597-3.

\bibitem{BFO-2} Y. Chu, M. Cruz, C. Yang, L. Martin, P. Yang, J.
Zhang, K. Lee, P. Yu, L. Chen, and R. Ramesh, Adv. Mater. \textbf{19}
(2007), 2662. https://doi.org/10.1002/adma.20060297231

\bibitem{BFO-3}S. H. Baek, H. W. Jang, C. M. Folkman, Y. L. Li, B.
Winchester, J. X. Zhang, Q. He, Y. H. Chu, C. T. Nelson, M. S. Rzchowski,
X. Q. Pan, R. Ramesh, L. Q. Chen, and C. B. Eom, Nat. Mater. \textbf{9}
(2010), 309. https://doi.org/10.1038/nmat2703

\bibitem{BFO-4}Z. Chen, Z. Chen, C.-Y. Kuo, L. W. Martin, et al.
Nat. Commun. \textbf{9} (2018), 3764. https://doi.org/10.1038/s41467-018-06190-5

\bibitem{CSI}I. Chung, J.-H. Song, J. Im, J. Androulakis, C. D. Malliakas,
H. Li, A. J. Freeman, J. T. Kenney, and M. G. Kanatzidis, J. Am. Chem.
Soc. \textbf{134} (2012), 8579. https://doi.org/10.1021/ja301539s

\bibitem{CSI-2}M. Liu, M. B. Johnston, and H. J. Snaith, Nature \textbf{501}
(2013), 395. https://doi.org/10.1038/nature12509

\bibitem{CSI-3}L.-y. Huang, and W. R. L. Lambrecht, Phys. Rev. B
\textbf{88} (2013), 165203. http://doi.org/10.1103/PhysRevB.88.165203

\bibitem{CSI-4}E. L. Da Silva, J. M. Skelton, S. C. Parker, and A.
Walsh, Phys. Rev. B \textbf{91} (2015), 144107. http://doi.org/10.1103/PhysRevB.91.144107

\bibitem{PBE}J. P. Perdew,  K. Burke, and M. Ernzerhof, Phys. Rev.
Lett. \textbf{77} (1996), 3865. https://doi.org/10.1103/PhysRevLett.77.3865

\bibitem{GPAW}P. E. Blöchl, Phys. Rev. B \textbf{50} (1994), 17953.
https://doi.org/10.1103/PhysRevB.50.17953

\bibitem{GPAW-2}J. Enkovaara, C. Rostgaard, J. J. Mortensen, et al.
J. Phys.: Condens. Matter \textbf{22} (2010), 253202. https://doi.org/10.1088/0953-8984/22/25/253202

\bibitem{pytilting}The program is at \url{https://gitlab.com/pyseries/pytilting}
and the documentation is at \url{https://dwang5.github.io/PyTiltingDoc/}.

\bibitem{LDA}R. E. Cohen, and H. Krakauer, Phys. Rev. B \textbf{42}
(1990), 6416. https://doi.org/10.1103/PhysRevB.42.6416

\bibitem{LDA-2}R. E. Cohen, Nature \textbf{358} (1992), 136. https://doi.org/10.1038/358136a0

\bibitem{M-P}H. J. Monkhorst, and J. D. Pack, Phys. Rev. B \textbf{13}
(1976), 5188. https://doi.org/10.1103/PhysRevB.13.5188

\bibitem{VESTA}K. Momma, and F. Izumi, J. Appl. Cryst. \textbf{44}
(2011), 1272. https://doi.org/10.1107/S0021889811038970

\bibitem{Phonony} A. Togo, L. Chaput, and I. Tanaka, Phys. Rev. B
\textbf{91} (2015), 094306. https://doi.org/10.1103/PhysRevB.91.094306

\bibitem{137}K. Leinenweber, and J. Parise, J. Solid State Chem.
\textbf{114} (1995), 277. https://doi.org/10.1006/jssc.1995.1040

\bibitem{CSI-5}L.-y. Huang, and W. R. L. Lambrecht, Phys. Rev. B
\textbf{90} (2014), 195201. https://doi.org/10.1103/PhysRevB.90.195201

\bibitem{B-F}N. A. Benedek, and C. J. Fennie, J. Phys. Chem. C \textbf{117}
(2013), 13339. https://doi.org/10.1021/jp402046t

\bibitem{iion}R. D. Shannon, Acta Cryst.: Sect. A \textbf{32} (1976),
751. https://doi.org/10.1107/S056773947600155
\end{thebibliography}
\end{document}